\documentclass[12pt]{iopart}
\usepackage{iopams}
\usepackage{epsf}
\usepackage{graphicx}
\usepackage{psfrag}

\newcommand{\beq}[0]{\begin{equation}}
\newcommand{\eeq}[0]{\end{equation}}

\def\Indic{\hbox{1\kern-.24em\hbox{I}}}

\begin{document}

\rightline{FNT/T 2004/12}

\title{Pricing Exotic Options in a Path Integral Approach}

\author{G Bormetti\dag\ddag, G Montagna\dag\ddag\footnote[4]
{Author to whom any correspondence should be addressed.}, 
N Moreni\S\dag\ and O Nicrosini\ddag\dag}

\address{\dag\ Dipartimento di Fisica Nucleare e Teorica, Universit\`a 
di Pavia, Via A. Bassi 6, 27100, Pavia, Italy}

\address{\ddag\ Istituto Nazionale di Fisica Nucleare, Sezione di Pavia, 
Via A. Bassi 6, 27100, Pavia, Italy} 

\address{\S\ CERMICS - ENPC, 6 et 8 avenue Blaise Pascal, Cit\'e Descartes, 
Champs sur Marne, 77455, Marne la Vall\'ee, Cedex 2, France }

\begin{abstract}
In the framework of Black-Scholes-Merton model of financial derivatives,
a path integral approach to option pricing is presented.
A general formula to price European path dependent options on multidimensional 
assets is obtained and implemented by means of 
various flexible and efficient algorithms.
As an example, we detail the cases of  Asian, barrier knock out, reverse cliquet
and basket call options, evaluating prices and Greeks. The numerical results 
are compared with those obtained with other procedures 
used in quantitative finance and found to be in good agreement. 
In particular, when pricing at-the-money and out-of-the-money options,
the path integral approach exhibits competitive performances.

\end{abstract}



\ead{giacomo.bormetti@pv.infn.it guido.montagna@pv.infn.it 
moreni@cermics.enpc.fr and oreste.nicrosini@pv.infn.it}

\maketitle

\section{Introduction and motivation}\label{s:intro}

A central problem in quantitative finance is the development of efficient 
methods for pricing and hedging derivative 
securities~\cite{Hull,Clewlow,Wilmott}. Although the 
classical Black, Scholes and Merton model of financial 
derivatives~\cite{bs,merton} provides an elegant 
framework to price financial derivatives, its actual 
analytical tractability is limited to plain vanilla call and 
put options and to few other cases. Actually, even if some particular payoffs 
lead to exact or approximated closed-form pricing 
formulae~\cite{Lo,Vecer}, these analytical results can be 
extended to more 
general payoffs only with difficulty. Hence, there is a need for flexible and fast pricing algorithms, 
especially when we are interested in
pricing options whose payoff at the expiry 
date depends on the whole path followed by the underlying 
(i.e.~path dependent options). In the past years many approaches have been 
proposed and the standard numerical procedures
adopted in financial engineering involve the use of binomial or 
trinomial trees,
Monte Carlo simulations and finite difference 
methods~\cite{Hull,Clewlow,Wilmott}. Alternative and
more recent algorithms are described, for example, in~\cite{Airoldi}, 
which has a comprehensive bibliography.

In this paper we extend the path integral approach to option pricing
developed for unidimensional assets in  \cite{Montagna}. 
We generalize the original formulation
in order to price a variety of commonly traded exotic options. 
First, we obtain a pricing formula for path dependent options 
based on multiple correlated underlying assets; second, we
improve the related numerical algorithms. 
Comparisons with standard Monte Carlo simulations, as well
as with the results of other numerical techniques known in the 
literature, are presented. Related attempts to price 
options and, in particular, exotic options, using the path integral 
method can be found 
in~\cite{Baaquie0,Linetsky,RCT0,Matacz0,RCT1,Baaquie1,Lyasoff,Dash}.

The structure of the article is the following. In 
Section \ref{s:path} we trace the derivation of the 
central pricing formula of our path integral-inspired 
approach and we describe in Section~\ref{s:algo} 
how to implement it numerically. Details about 
the computational algorithms used for pricing and numerical 
results are discussed in Section \ref{s:numerical}. 
In this latter, we show that our
approach can be efficiently implemented to price a 
large class of exotic options: 
Asian, barrier knock out and  reverse cliquet. 
Finally, in Section \ref{s:greek} we compute the 
Greeks relative to some of the considered options and 
in Section \ref{s:conclusion} we draw conclusions and 
consider possible perspectives.


\section{A path integral-based pricing formula}
\label{s:path}
Path integral techniques, 
 widely used in quantum mechanics and quantum field theory, 
can be useful to describe the dynamics of a Markov stochastic 
process~\cite{FeynmanHibbs,Schulman,Chaichian}. In 
the present paper we are interested  in computing 
mean values of functionals of a $D$-dimensional Markov 
stochastic process $S(t)$ corresponding to the price of a 
set of $D$ underlying assets.

In particular, let us fix a time horizon $T>0$ and 
split the time interval $[0,T]$ into $n+1$ subintervals 
$[T_i,T_{i+1}]$ with $T_0\equiv 0$, 
$T_{n+1}\equiv T$ and $T_{i+1}-T_i\doteq\Delta t \doteq T/(n+1)$. 
Our aim is to compute the  fair price  at time $T_0$ of 
a European path dependent option with maturity $T$ and 
whose payoff $f$ is a function of the values $S(0),S(T_1),\ldots,S(T_{n+1})$.

 According to the arbitrage-free pricing theory 
 \cite{bs,merton,Bjork}, this means that we have to  
evaluate the mathematical expectation\footnote{For simplicity, 
we have included the discount factor $\exp\{-rT\}$ in the definition 
of the payoff.}
\begin{eqnarray}\label{targetExpectation}
 \fl\mathbb{E}[f(Z(T_{n+1}),Z(T_n),\ldots,Z_0)]=\nonumber\\
=\int_{\mathbb{R}^{D\times(n+1)}}\hspace{-1cm}\rmd z_{n+1}
\cdots\rmd z_1~p(z_{n+1},z_n,\ldots,z_0)
 f(z_{n+1},z_n,\ldots,z_0),
\end{eqnarray}
where $Z(t)\doteq \log S(t)$ and  $p(z_{n+1},z_n,\ldots,z_0)$ is 
the joint probability density function (pdf) of the path $\{Z(T_0)=Z_0, 
\ldots , Z(T_i)=z_i , \ldots,  Z(T_{n+1})=z_{n+1}\}$. 
In order to compute Equation~\eref{targetExpectation}, 
a straightforward application of standard Monte Carlo estimation 
theory, would require the sampling of $N$ independent and identically 
distributed (i.i.d.) paths  $\{Z_0,Z^{(l)}(T_1),\ldots,Z^{(l)}(T_{n+1})\}_{l=1,\ldots,N}$. 
Whenever $S(t)$ (and hence $Z(t)$) is solution of a stochastic 
differential equation (SDE), this can be done by an 
Euler discretization scheme of the SDE. Unfortunately, 
this procedure may be slow, as to 
compute $Z^{(l)}(T_i)$  we typically need to know $Z^{(l)}(T_{i-1})$, and is not 
efficient when considering out-of-the-money (OTM) options, 
because a relevant number of sampled paths may not 
contribute to the payoff. That is why we look for an alternative 
formulation  of this pricing problem leading to a reduction of the 
Monte Carlo variance\footnote{We refer to~\cite{Clewlow,Glasserman} 
for a review of standard Monte Carlo variance reduction techniques.}.

\subsection{The model}

Let us first of all introduce the evolution model we will focus on throughout 
the rest of the paper. We assume that $S(t)$ satisfies, under 
the objective probability measure, the following SDE   
\beq \label{multisde}
\begin{array}{rcl}
dS^{k}(t)/S^{k}(t)&=&\mu^{k}dt+\sigma_{k}d\bar W^{k}(t)
\quad\qquad\forall k=1,\ldots,D\\
\langle d\bar W^{i}(t),d\bar W^{j}(t)\rangle&=&\rho_{ij}dt\qquad\qquad\qquad\qquad 
\forall i,j=1,\ldots,D,
\end{array}
\eeq
where the $\mu^k\in\mathbb{R}$ and the $\sigma_{k}\in\mathbb{R}^+$ 
represent the mean returns  and the volatilities of $S^k$, respectively, 
and the $\rho_{ij}\in (-1,1)$ are the correlations between the 
components of the Wiener processes $\bar W(t)$ ($\rho_{ii}=1$).
This formulation is particularly useful because it only involves 
financial quantities that can be historically estimated. For instance, 
$\rho_{ij}$ and $\sigma^{k}$ can be evaluated by analyzing the time series 
of the correlations between different assets' returns, i.e.
\begin{equation}
\begin{array}{rcl}
\langle dS^{i}(t),dS^{i}(t)\rangle&=&(S^{i}(t)\sigma_{i})^{2}dt\\
\langle dS^{i}(t),dS^{j}(t)\rangle&=&S^{i}(t)S^{j}(t)\sigma_{i}
\sigma_{j}\rho_{i,j}dt\quad i\neq j.
\end{array}
\end{equation}

However, it is convenient to write Equation~\eref{multisde} 
in terms of the square root $\Sigma$ of the variance-covariance 
matrix $\bar \Sigma_{i,j}\doteq \sigma_i\sigma_j\rho_{i,j}$ and of a standard 
$D$-dimensional Wiener process $W$. The square root $\Sigma$ is 
defined by relation $\Sigma\Sigma^T=\bar{\Sigma}$ 
and can be chosen to be a lower triangular matrix. 
Changing the probability measure from the original objective measure
of Equation (\ref{multisde}) to the risk neutral one \cite{Bjork}, the stochastic process 
$Z(t)\doteq(\log S_{1}(t),\ldots,\log S_{D}(t))$ satisfies the 
following equation
\beq \label{logsde}\left\{\begin{array}{l}
dZ(t)=Adt+\Sigma dW(t)\\
Z(0)=Z_0,
\end{array}\right.\eeq
where the $k^{th}$ entry of $A$ is $A^k=(r-\sigma^{2}_k/2)$, 
with $r$ the risk-free interest rate. From Equation \eref{logsde}, we infer  
that $Z(t)$ is normally distributed with mean $Z_0+At$ and 
variance-covariance matrix $\bar{\Sigma}t$. Equivalently, the 
conditional pdf~\footnote{By definition, $p(z', t'|z,t)$  
is such that the probability for $Z(t')$ taking a value 
in the $D$-dimensional hyper-cube $dz'$ centred on $z'$, 
conditional on $Z(t)=z$, is  $p(z', t'|z,t)dz'$.}
$p(z', t'|z,t)$, $t'>t$, is given by
\beq\label{lognorm}
 \fl p(z', t'|z,t)=\left(\frac{1}{2\pi (t'-t)}\right)^{D/2}
 \frac{1}{|\mathrm{det}\Sigma|}\exp{\left\{-\frac{1}{2(t'-t)} 
 ||\Sigma^{-1}(z'-z-A(t'-t))||^{2}\right\}},
\eeq
where $||\cdot ||$ stands for the standard Euclidean norm.
Solutions of Equation~\eref{logsde} are Markov processes and, therefore, 
it is possible to describe their 
time evolution via a path integral formulation~\cite{Montagna}.
 
\subsection{The fundamental pricing formula}

Thanks to the properties of the chosen model, we are now able 
to extend the pricing formula given in 
\cite{Montagna} and propose improved algorithms to evaluate 
Equation~\eref{targetExpectation}. Our approach is 
essentially based on a sequence of linear changes of the 
integration variables appearing in Equation~\eref{targetExpectation}. This 
latter expression will then be rewritten in terms of a  suitable set of 
independent random variables.

The definition of conditional probability, 
together with the Markov nature of the price dynamics, allows us to write
the joint probability entering Equation~\eref{targetExpectation} as 
\begin{eqnarray}\label{markovpdf}
p(z_{n+1},z_n,\ldots,z_0)\prod_{i=1}^{n+1}\rmd z_i
&=&\prod_{i=1}^{n+1}\rmd z_i~p_{i-1}(z_i|z_{i-1})\nonumber \\ 
&=&\prod_{i=1}^{n+1}\rmd z_i\left[\left(\frac{1}{2\pi\Delta t}
\right)^{D/2}\!\!\! \frac{1}{|\mathrm{det}\Sigma|}
\rme^{-||\Sigma^{-1}[z_{i}-(z_{i-1}+A\Delta t)]||^{2}/2\Delta t}
\right],\nonumber\\ &&
\end{eqnarray}
where we have written the transition densities 
$p_{i-1}(z_i|z_{i-1})\doteq p(z_i,T_i|z_{i-1},T_{i-1})$ explicitly.
 In order to get rid of the correlation matrix $\Sigma$ and 
 of the drift $A$, we perform a first change of variable by 
 setting $z_{i}=\Sigma(\eta_{i}+Ai\Delta t)$, $i=0$ to $n+1$, thus obtaining
 for the r.h.s. of Equation (\ref{markovpdf})
\begin{equation}\label{pdfEta}
 \mathrm{r.h.s.~(\ref{markovpdf})}=\prod_{i=1}^{n+1}\left
 [\left(\frac{1}{2\pi\Delta t}\right)^{D/2}\rmd \eta_{i}
 \exp{\left\{-\frac{1}{2\Delta t}||\eta_{i}-\eta_{i-1}||^{2}\right\}}\right].
\end{equation}

We work out the quadratic form 
$\sum_{i=1}^{n+1}||\eta_i-\eta_{i-1}||^2$ and rearrange 
terms by introducing the $D$-dimensional vectors $h_1,\ldots,h_{n}$ such that 
$$\eta^{k}_{i}=\sum_{j=1}^{n} O_{ij}h^{k}_{j}\quad 
 k=1,\ldots D,\; i=1,\ldots,n,$$ 
 where  $\bi{O}$ is the orthogonal matrix that diagonalizes 
 the $n\times n$ tridiagonal matrix
\beq
 \bi{M}=\left( \begin{array}{llllll}
 2&-1&0&\cdots&\cdots&0 \\
 -1&\ 2&-1&0&\cdots&0 \\
 0&-1&\ 2&-1&\cdots&0\\
 0&\cdots&-1&\ 2&-1&0 \\
 0&\cdots&\cdots&-1&\ 2&-1\\
 0&\cdots&\cdots&\cdots&-1&\ 2 \end{array} \right).
\eeq
After some tedious, but straightforward, algebra, we obtain
\begin{eqnarray}\label{pdfH}
\mathrm{r.h.s.~(\ref{pdfEta})}=g(z_{n+1};z_0)
\rmd z_{n+1}\prod_{i=1}^{n}\rmd h_{i}\varrho_{i}(h_{i};z_0,z_{n+1}),
\end{eqnarray}
where  the $\varrho_{i}(\;\cdot\; ;z_0,z_{{n+1}})$ are $D$-dimensional 
Gaussian pdfs with mean 
$\vartheta_i=[\Sigma^{-1}z_{0}O_{1i}+\Sigma^{-1}(z_{n+1}-Ai\Delta t)O_{ni}]/m_i$
 and variance $(\Delta t/m_i)\Indic_{D\times D} $, the $\{m_i\}_{i=1,\ldots,n}$
  being the eigenvalues of $\bi{M}$. The function $g$ is defined as 
\begin{eqnarray*}
\fl g(z_{n+1};z_0)=\frac{1}{|\mathrm{det}\Sigma|}\left(\frac{1}{\sqrt{2\pi\Delta t \mathrm{det}(\bi{M})}}\right)^{D} \!\!\!\!\! \\
\hspace{1.5cm} \exp 
\left\{\frac{ ||\Sigma^{-1} z_{0}||^{2}+||\Sigma^{-1}
(z_{n+1}-A(n+1)\Delta t)||^{2}-\sum_{i=1}^{n} ||\vartheta_i||^{2}/m_{i}}
{2\Delta t }\right\},
\end{eqnarray*}
and from now on we will drop, for simplicity, its dependence on $z_0$.

Finally, we replace the $h_i$ by setting 
$h_i=\vartheta_i+\lambda_i\sqrt{\Delta t/m_i}$, thus obtaining the ultimate 
relationship
\begin{eqnarray}\label{pdfLambda}
\mathrm{r.h.s.~(\ref{pdfH})}=g(z_{n+1})
\rmd z_{n+1}\prod_{i=1}^{n}\rmd \lambda_{i}\rho_{G}(\lambda_{i}),
\end{eqnarray}
where $\rho_G$ is a $D$-dimensional standardized Gaussian pdf.
By means of this sequence of replacements, 
the expectation \eref{targetExpectation} can be computed as
\begin{eqnarray}\label{targetIO}
 \fl\mathbb{E}[f(Z(T_{n+1}),\ldots,Z_0)]=\nonumber \\
=\int_{\mathbb{R}^{D}}\!\!\rmd z_{n+1}g(z_{{n+1}})
\int_{\mathbb{R}^{D\times n}}\prod_{i=1}^{n}\left(\rmd \lambda_{i}
\rho_G(\lambda_{i})\right)\tilde f(z_{n+1},\lambda_n,\ldots,\lambda_1,z_0)
\end{eqnarray}
where $\tilde f(z_{n+1},
\lambda_n,\ldots,\lambda_1,z_0)\doteq f(z_{n+1},z_n,\ldots,z_1,z_0)$ 
with the substitution
\begin{equation}\label{pathconstruct}
z_{i}=\sum_{j=1}^{n} O_{ij}\Sigma\left(\sqrt{\frac{\Delta t}{m_i}}
\lambda_{j}+\vartheta_j\right)+iA\Delta t, \quad  i=1,\ldots,n.
\end{equation}  

The reformulation of Equation~\eref{targetExpectation} as in 
Equation \eref{targetIO}  is the core of our pricing technique. 
In particular, advantages come from having split 
the $D\times (n+1)$-dimensional integral into an \emph{external} 
integration over $z_{n+1}$, representing the value of the log-price 
at the maturity,  and  an \emph{internal} one, which can be 
thought as the mathematical expectation 
\begin{equation*}
\mathbb{E}[\tilde f(z_{n+1},\Lambda_n,\ldots,\Lambda_1,z_0)]
=\int_{\mathbb{R}^{D\times n}}\prod_{i=1}^{n}
\left(\rmd \lambda_{i}\rho_G(\lambda_{i})\right)
\tilde f(z_{n+1},\lambda_n,\ldots,\lambda_1,z_0),
\end{equation*}
where  $\Lambda_1,\ldots,\Lambda_n$ are standardized 
$D$-dimensional i.i.d.~Gaussian  variables.

 Let us stress that, for each value of $z_{n+1}$, and since $z_0$ is known, 
 by means of $N$  i.i.d.~Gaussian samples $\left\{(\Lambda_1^{(l)},
 \ldots,\Lambda_n^{(l)})\right\}_{l=1,\ldots,N}$, 
 we can construct the set of log-price paths 
\begin{equation}\label{logPath}
Z^{(l)}(T_{i})=\sum_{j=1}^{n} O_{ij}
\Sigma\left(\sqrt{\frac{\Delta t}{m_i}}
\Lambda^{(l)}_{j}+\vartheta_j\right)+iA\Delta t,
\end{equation}
having  fixed starting and end
points $Z^{(l)}(T_{n+1})=z_{n+1}$ and $Z^{(l)}(T_0)=z_0$.  
This way of proceeding is typical of path integral 
techniques, in which functional trajectories 
with fixed  initial and final states are considered 
\cite{FeynmanHibbs,Schulman,Chaichian}. That is why we call our 
method \emph{path integral pricing}.

\section{Computational algorithms}\label{s:algo}

The reformulation of the pricing problem given in 
Section~\ref{s:path}  is  useful to price path dependent options: 
this task has been reduced to the numerical computation of the integrals 
appearing in Equation~\eref{targetIO}. In particular, we can adopt 
the two following procedures
\begin{itemize}
 \item[1.] We can compute the internal $D\times n$-dimensional 
 integral via Monte Carlo sampling of the 
 $\Lambda_i$, and the external $D$-dimensional one  
 by a deterministic method to be specified. 
 We will call this method \emph{path integral with external integration}. 
 This method turns out to perform well when $D=1$, as shown
 in the following.
 \item[2.] We  can perform a pure $D\times (n+1)$-dimensional 
 Monte Carlo integration by properly truncating the integration 
 domain on $z_{n+1}$. 
 This method will be called \emph{pure Monte Carlo} and is 
 particularly useful when considering OTM options on multidimensional assets.
\end{itemize}

We provide, in the next two subsections, 
a more exhaustive  insight into the  procedures sketched above. 
We refer to Section \ref{s:numerical} for the implemented versions' 
details and the numerical results. 

\subsection{Path integral with external integration}\label{pathextsec}
This method corresponds to a very precise evaluation 
of the inner function 
$\mathbb{E}[\tilde f(z_{n+1},\Lambda_n,\ldots,\Lambda_1,z_0)]\doteq
\mathcal{E}(z_{n+1})$ for some given values of $z_{n+1}$. 
Actually, our aim is to approximate Equation~\eref{targetIO} by a formula like
\beq\label{GQuad}
\int_{\mathbb{R}^{D}}\!\!\!\!\!\rmd 
z_{n+1}~g(z_{{n+1}})\mathcal{E}(z_{n+1})
\approx\sum_{i=1}^{n_{int}}g(z^{(i)}_{{n+1}})\mathcal{E}(z^{(i)}_{n+1})w_{i},
\eeq 
with a suitable choice of the integration weights $w_{i}$ and of the 
integration  points $z_{n+1}^{(i)}$. We can, for example, perform Riemann 
integration or exploit a quadrature rule \cite{numericalC}. Since 
$\mathcal{E}(z^{(i)}_{n+1})$ is a non-explicitly solvable 
mathematical expectation, for each $z^{(i)}_{n+1}$ we estimate it 
by sampling $N$ i.i.d.~from the law of $(\Lambda_1,\ldots,\Lambda_n)$, 
thus obtaining, at the same time, the associated errors $v_i$. By virtue 
of the Central Limit Theorem, the $v_i$ scale with the square root of $N$, 
so that the bigger $N$ is, the smaller the error and more precise are the 
values of $\mathcal{E}(z^{(i)}_{n+1})$. Of 
course, the choice of the $z_{n+1}^{i}$ influences the 
final result and has to be done carefully. By means of these 
coupled outer-deterministic and inner-Monte 
 Carlo integrations, we estimate the price with
\beq\label{meanerrorbar}
 B_{\pm}\doteq\sum_{i=1}^{n_{int}}w_i g(z^{(i)}_{{n+1}})
 \mathcal{E}(z_{n+1}^{(i)})
 \pm \sqrt{\sum_{i=1}^{n_{int}}(g(z^{(i)}_{{n+1}}) w_i v_i)^2},
\eeq
as boundary values for the $68\%$ confidence 
interval of Equation~\eref{targetIO}.
It is worth noticing that such an error does not include the effect 
of finiteness of $n_{int}$. Numerical results providing 
us with evidence the error due to a finite $n_{int}$ 
is negligible are reported in Section \ref{s:asianunidim}.

To conclude, let us remark that this  procedure, 
forcing $Z(T_{n+1})$ to take a value 
in $\{ z^1_{n+1},\ldots,z^{n_{int}}_{n+1}\}$, is similar to the 
 variance reduction technique known as 
 \emph{stratified sampling Monte Carlo} \cite{Glasserman,Lapeyre}. 

\subsection{Pure Monte Carlo}\label{s:MCpuresec}

We will show in the next section that when pricing unidimensional 
assets according to Equation~\eref{meanerrorbar}, a deterministic 
choice of final integration points works better than a Monte Carlo one. 
However, it is known that  deterministic integration 
approaches rapidly lose their competitiveness as the dimension grows. 
As an alternative, we propose a method based on a pure Monte Carlo integration  coupled with the path integral.\\
First, we choose a function $\Gamma:\mathbb{R}^D\rightarrow (0,+\infty)$ 
such that $\Gamma > 0$ and $\int_{\mathbb{R}^D}\Gamma(z)dz=1$ and 
we interpret it as a pdf.
Second, we rewrite Equation~\eref{targetIO} as 
\begin{eqnarray}\label{targetMC}
 \fl\mathbb{E}[f(Z(T_{n+1}),\ldots,Z_0)]&=&
 \int_{\mathbb{R}^{D\times(n+1)}}\!\!\rmd z_{n+1}
 \Gamma(z_{n+1})\prod_{i=1}^n\rho_G(\lambda_i)
 \rmd\lambda_i \hat f(z_{n+1},\lambda_n,\ldots,\lambda_1,z_0)\nonumber \\ 
&=&\mathbb{E}[\hat f(Z,\Lambda_n,\ldots,\Lambda_1,z_0)],
\end{eqnarray} 
where $\hat f(z_{n+1},\lambda_n,\ldots,\lambda_1,z_0)\doteq g(z_{{n+1}})
\tilde f(z_{n+1},\lambda_n,\ldots,\lambda_1,z_0)/\Gamma(z_{n+1})$ and $Z$ 
is a random variable with $\Gamma$ as pdf. In other words, 
we read the pricing formula 
as the mathematical expectation of a function of $n+1$ independent
 variables, namely $Z$ and the $\Lambda_i$. Our algorithm evaluates 
 Equation~\eref{targetMC} by a pure Monte Carlo method
 extracting $N$ random i.i.d.~samples 
 $(Z^{(l)},\Lambda^{(l)}_1,\ldots,\Lambda^{(l)}_n)_{l=1,\ldots,N}$. 

This method resembles a standard Monte Carlo simulation of 
random walks, but there are some subtle differences. 
First of all, in the random walk case one simulates each path 
recursively by sampling $n+1$ Gaussian variables without 
knowing where the considered path will end, while here we want 
to construct paths leading to a given $z_{n+1}$. Second, we 
introduce an asymmetry between $z_{n+1}$ and the 
$\lambda_i$ in the sense that $z_{n+1}$ 
plays a crucial role and we give to it the possibility of being 
sampled by a non-Gaussian pdf by changing $\Gamma$. 
This turns out to be very useful when 
  pricing OTM options and the Monte Carlo random walk turns out to be 
  not efficient, as shown in the next section.    


\section{Numerical results and discussion}\label{s:numerical}

In this section we 
report computational issues concerning the pricing of different
kinds of path dependent options by means of the 
path integral procedures discussed in Section~\ref{s:algo}. 
We will consider the following types of 
options: Asian and up-out barrier unidimensional 
call, unidimensional reverse cliquet and Asian basket call. 
The dynamics of the underlying assets is 
assumed to follow Equation \eref{logsde}. \par
\subsection{Algorithms' implemented  versions}\label{s:listofalgo}
Before entering into details with numerical results, let us list 
here which  versions of the two general algorithms of 
Section \ref{s:algo} have been implemented.
\begin{enumerate}
\item \emph{Path Integral with Trapezoidal Integration} (PITP)

This is an algorithm of the type described in Section \ref{pathextsec}. 
In particular, as deterministic method used to integrate over $z_{n+1}$, 
we choose trapezoidal integration with equispaced abscissa 
\cite{numericalC}. Integration, whenever not differently 
specified, instead of being performed 
on $\mathbb{R}^D$, is truncated, for each $k=1$ to $D$, to the 
interval 
$$\mathcal{C}_k\doteq [\bar z^k -4\sigma_k\sqrt{T},\bar 
z^k +4\sigma_k\sqrt{T} ],$$ 
where $\bar z^k \doteq Z^k_0+(r-\sigma_k^2/2)T$ 
for in-the-money (ITM) and at-the-money (ATM) options and 
$\bar z=\log(K)$ for  OTM ones. Tests regarding this choice 
can be found in Section \ref{s:asianunidim}. 

\item \emph{Pure Monte Carlo with Flat Sampling} (PIFL)

This is a Pure Monte Carlo algorithm (see Section \ref{s:MCpuresec}) 
in which we sample the $Z{(l)}$ of Equation \eref{targetMC} from a uniform 
pdf on the compact subset 
$\mathcal{C}\doteq \mathcal{C}_1\otimes\cdots\otimes \mathcal{C}_D$.

\item \emph{Pure Monte Carlo with Truncated Cauchy Sampling} (PICH)

  This version of the Pure Monte Carlo method uses, as $\Gamma$, 
   a truncated Cauchy pdf centred on $\bar z$ and normalized to 
   one on the compact subset $\mathcal{C}$. The particular choice of a Cauchy 
   function is suggested by a simple heuristic reasoning and confirmed 
   by empirical tests. Let $f$ be the payoff of a vanilla 
   (call or put) option with $D=1$. In order to compute 
   Equation~\eref{targetExpectation}, we should  integrate 
   a function that is of the form of the product of a max$(\cdot,\cdot)$ with 
   a Gaussian pdf. As a consequence, there will be 
   a region $z_{n+1}\in (-\infty,z_{low}]\cup [z_{up},+\infty)$ 
   in which the integrand is (almost) zero and a region in which 
   we expect it to be slightly  wider than a Gaussian one. We will 
   come back to this point at the beginning of Section 
   \ref{s:asianunidim}.
   
\end{enumerate}

As global  benchmarks, we will quote results obtained by means of a 
standard Monte Carlo random walk (MCRW) technique. 
In other words, we  sample $N$ i.i.d.~paths 
$\{Z_0,Z^{(l)}(T_1),\ldots,Z^{(l)}(T_{n+1})\}_{l=1,\ldots,N}$, 
built up by iterating, for all $i=1$ to $n+1$
\begin{equation}\label{logEuler}
Z^{(l)}(T_i)=Z^{(l)}(T_{i-1})+A\Delta t 
+\sqrt{\Delta t}\Sigma \xi^{(l)}_i\quad i=1,\ldots,N,
\end{equation}
where the $\xi^{(l)}_i$ are  i.i.d.~standardized $D$-dimensional Gaussian random variables.

Furthermore, in order to compare the PITP algorithm with a 
method similar in spirit, we implemented, in the case  $D=1$, a 
stratification-like  algorithm based on L\'evy recursive 
construction of Brownian motion, the so called Brownian bridge.
Details about this testing algorithm, that we will refer to 
as the  Brownian bridge with stratification (BBST), are reported, 
for completeness, in~\ref{s:bbridge}. 
In some cases, the algorithms are 
implemented by doubling the number of paths 
using antithetic variance reduction~\cite{Clewlow}. 
Whenever this is done, the corresponding algorithm 
is pre-fixed by the letters AT.    

\subsection{Unidimensional asset}\label{s:uni}
\subsubsection{Asian option\\}\label{s:asianunidim}

The fair price for a discretely sampled Asian call option 
on an unidimensional asset is   
\beq\label{asian}
\begin{array}{rcl}
\mathcal{O}_A(Z_0)&=& \mathbb{E}[f_A(Z(T_{n+1}),\ldots,Z_0)]\\
f_A(Z(T_{n+1}),\ldots,Z_0)&=&~\rme^{-rT}\max 
\left\{\left(\sum_{i=0}^{n+1}\exp\{Z(T_i)\}\right)/(n+2)-K, 0\right\},
\end{array}
\eeq
where $K$ is the strike price and $T$ the maturity. 
The parameters used in the numerical simulation are: $Z_0=\log 100$, $r=0.095$, $\sigma=0.2$, $t=0$, $T=1$ year and 
$n+1=100$. For simplicity, we omit the labels in the definition of 
$Z_0$ and $\sigma$ for all the unidimensional assets. 
Moreover, we consider $K=60,100,150$ in order to test the 
performances of our algorithm when the option is 
ITM, ATM and OTM, respectively.
    
\begin{figure}[t!]
 \caption{\label{integrand} Shape of the integrand function 
 $g(z_{n+1})\mathcal{E}(z_{n+1})$
 of Equation \eref{targetIO} for an Asian call option, showing 
 how the support and the  value of the maximum change when 
 considering in-the-money (top left), 
 at-the-money (top right) and out-of-the-money (bottom left) options.}
 \begin{minipage}[b]{0.55\textwidth}
  \begin{center}
   \psfrag{z_{n+1}}{\footnotesize $z_{n+1}$}
   \psfrag{I[z_{n+1}]}{\footnotesize $g(z_{n+1})\mathcal{E}(z_{n+1})$} 
   \psfrag{100}{\footnotesize $100$}
   \includegraphics[width=8.5cm]{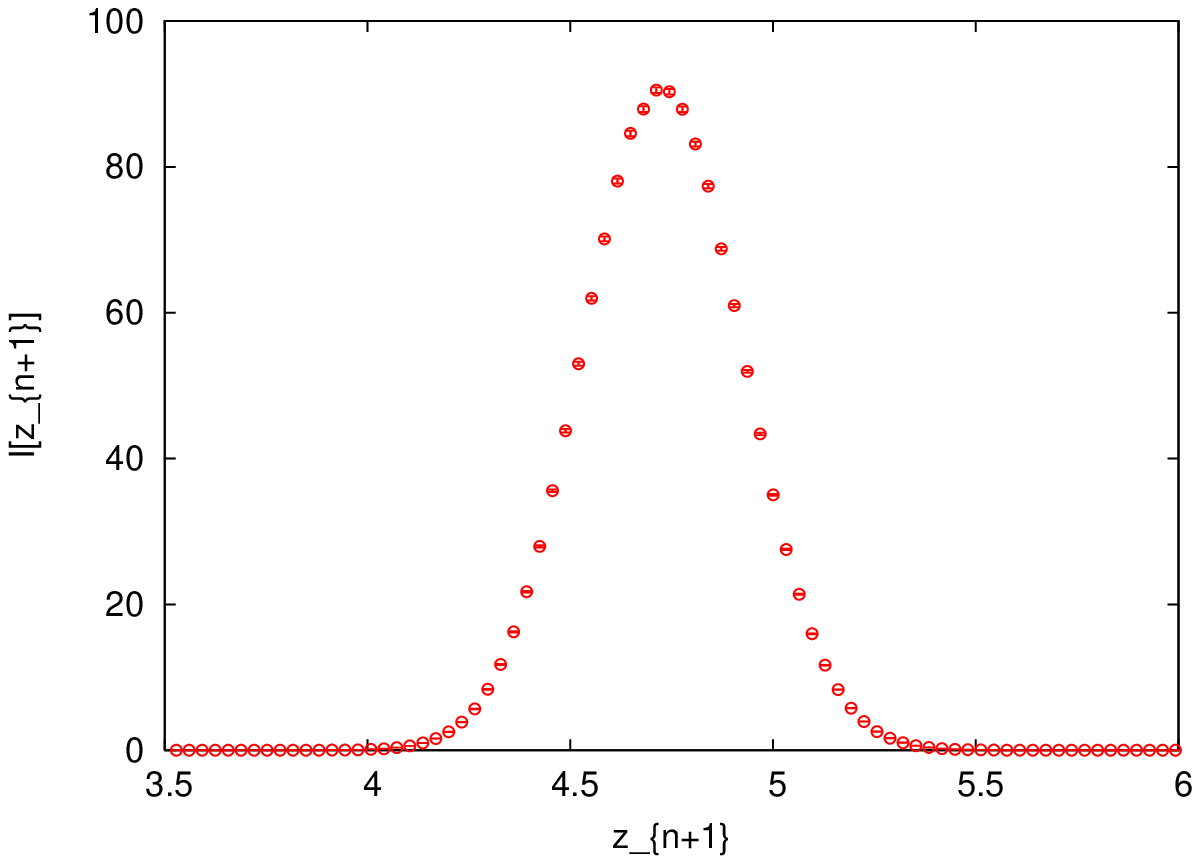}
  \end{center}
 \end{minipage}
 \begin{minipage}[b]{0.5\textwidth}
  \begin{center}
   \psfrag{z_{n+1}}{\footnotesize $z_{n+1}$}
   \psfrag{I[z_{n+1}]}{\footnotesize $g(z_{n+1})\mathcal{E}(z_{n+1})$}
   \includegraphics[width=8.5cm]{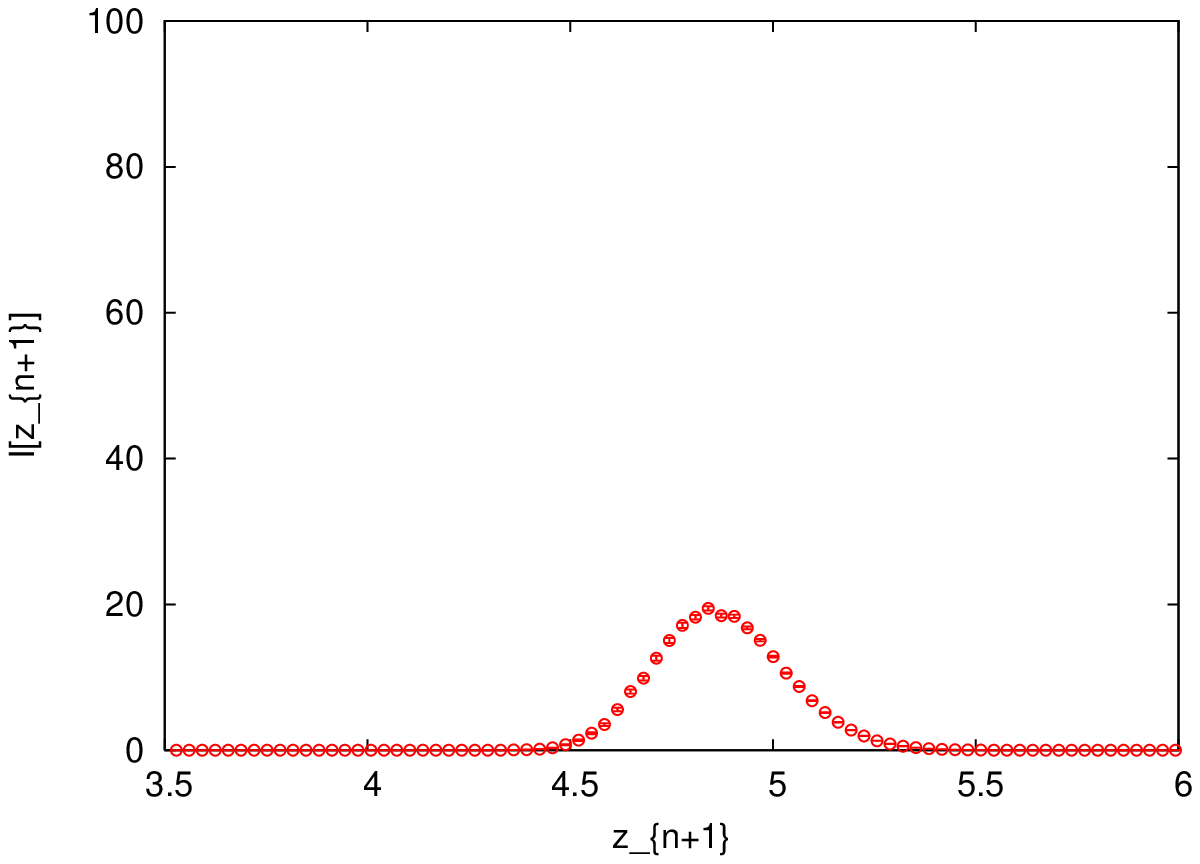}
  \end{center}
  \end{minipage}
\end{figure}

We justify the choices made about the integration domain 
discussed in Section \ref{s:listofalgo} as follows. Let us approximatively trace 
the shape of the integrand function $g(z_{{n+1}})\mathcal{E}(z_{n+1})$ 
in Equation~\eref{targetIO} for   Asian 
call options. In Figure \ref{integrand} we show the results obtained 
for an ATM, an ITM and an OTM option. Error bars 
correspond to one standard deviation Monte Carlo errors.
\begin{figure}[ht!]
 \begin{minipage}[b]{0.55\textwidth}
 \begin{center}
   \psfrag{z_{n+1}}{\footnotesize $z_{n+1}$}
   \psfrag{I[z_{n+1}]}{\footnotesize $g(z_{n+1})\mathcal{E}(z_{n+1})$}
   \includegraphics[width=8.5cm]{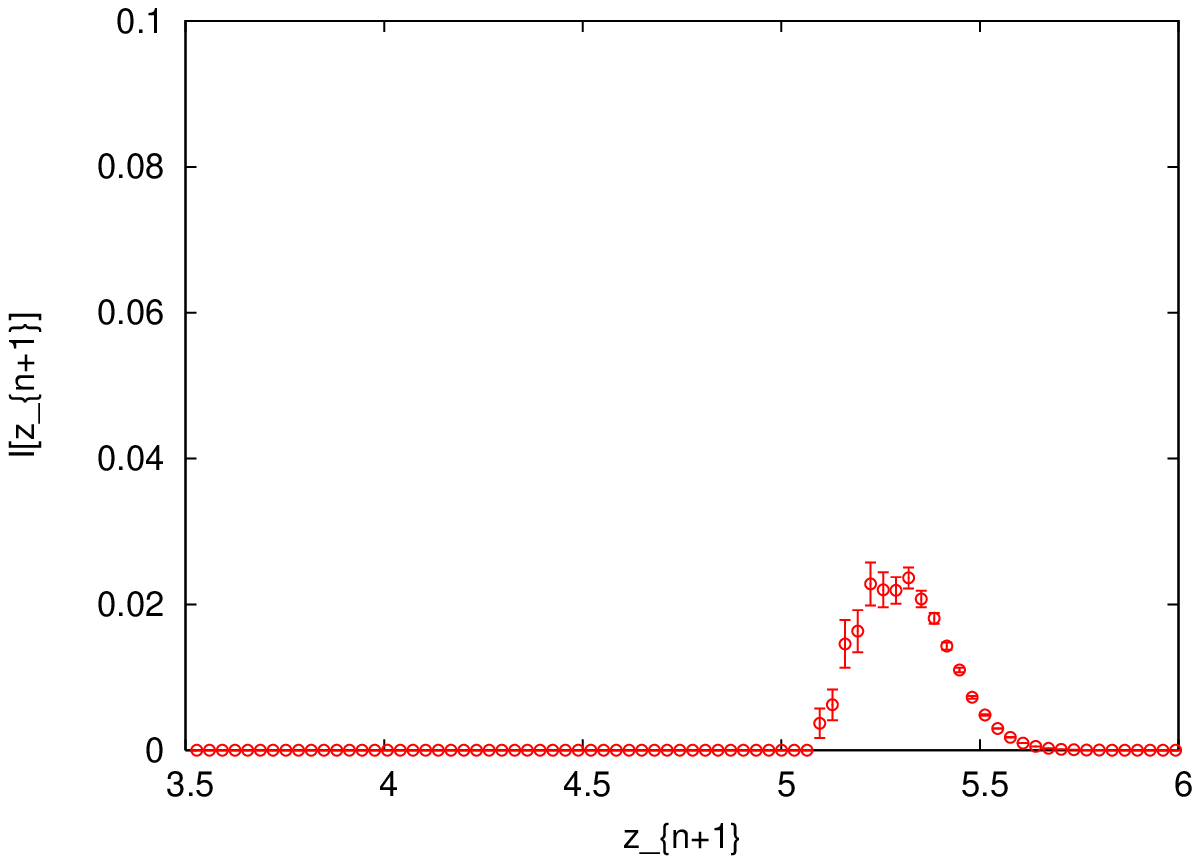}
  \end{center}
 \end{minipage}
 \begin{minipage}[b]{0.5\textwidth}
 \end{minipage}
\end{figure}
It is particularly interesting to study the support of the integrand 
function. Actually, for ITM and ATM options, the values of $z_{n+1}$ 
for which the function is considerably different from zero are more 
or less centred at $Z_0+(r-\sigma^2/2)T$, as 
can be seen from Figure \ref{integrand}. 
On the other hand, for OTM options, the lower bound 
is $\gtrapprox\log K$. Hence, we can exploit this property as a rule of 
thumb to reduce the external integration  to a (small) interval significantly 
contributing to the integral and 
to eventually perform importance sampling with an appropriate pdf. 
That is why we adopt particular values for the centre $\bar z$
of the integration domain.

Let us now come to the discussion of the 
numerical results. In Table \ref{asiantable} we report 
option prices and relative numerical errors as 
obtained by means of our path integral-based algorithms, as well as 
by the benchmarks. In the PITP and BBST cases, 
the number of integration points $n_{int}$ is set to 200 and for each 
point we generate $N=1000$ random paths.
In the cases of MCRW and of pure Monte Carlo path integral with flat 
(PIFL) or Cauchy (PICH) sampling, the total number of paths is 
$2\times10^5$, so that we compare results obtained 
with the same number of random calls and the comparison is meaningful.
In the lower part of the table, we present the results 
of some of the algorithms improved by the implementation of the 
antithetic variables technique. 
We notice that all path integral prices are in very good agreement with the 
ones obtained via random walks and BBST. As a further 
cross-check, we compared our path integral predictions with the  
results of the method developed in~\cite{Airoldi}, finding perfect agreement. 
>From the point of view of variance reduction, the PITP algorithm turns 
out to be the method that performs best, 
especially when pricing ATM/OTM options.  This means that, 
when the integrand is non-zero only in a region far from 
$Z_0+(r-\sigma^2/2)T$, the standard MCRW generates many paths that are not relevant for 
the mathematical expectation. Furthermore, the PITP and the 
BBST techniques give essentially the same results, thus confirming 
our suspicion that the strategy of fixing the end point before generating paths
plays the crucial role in the variance reduction. Let us stress 
that the flat integration gives a worse variance and that PICH 
and PIFL algorithms perform best only out of the money.

%
%
%
\Table{\label{asiantable} Numerical values for an Asian 
call option price obtained via different algorithms for the 
parameters $S_0=100$, $r=0.095$, $\sigma=0.2$, $T=1$ year and $n+1=100$. 
Errors correspond to one standard deviation.}
\br
& \centre{2}{ITM$^{\rm a}$} & \centre{2}{ATM$^{\rm b }$} & 
\centre{2}{OTM$^{\rm c}$} \\
\ns
&\crule{2}&\crule{2}&\crule{2}\\
& Price & Error & Price & Error & Price & Error\\
\mr 
MCRW &40.830&0.025& 6.899 & 0.019 &0.0054&0.0005\\
BBST &40.824&0.018& 6.886 & 0.015 &0.0058&0.0001\\
PITP &40.811&0.019& 6.876 & 0.015 &0.0057&0.0001\\
PICH &40.767&0.040& 6.873 & 0.019 &0.0059&0.0001\\
PIFL &40.758&0.105& 6.880 & 0.026 &0.0057&0.0001\\
\mr
AT-MCRW &40.836&0.002&6.909&0.008&0.0053&0.0003\\
AT-PITP &40.832&0.004&6.901&0.004&0.0060&0.0001\\
AT-PICH &40.775&0.031&6.878&0.008&0.0058&0.0001\\
\br
\end{tabular}
\item[]$^{\rm a}$ In-the-money, $K=60$.
\item[]$^{\rm b}$ At-the-money, $K=100$.
\item[]$^{\rm c}$ Out-of-the-money, $K=150$.
\end{indented}
\end{table}
%

\begin{figure}[t!]
 \caption{\label{f:error} Prices and error bars for an at-the-money Asian 
 call option as a function of the number of points for external
 integration $n_{int}$. As $n_{int}$ increases, the error 
 bars decrease ($\sim n_{int}^{-0.5}$), while the fluctuation 
 of prices reduces.}
 \centering
 \psfrag{n_{int}}{\footnotesize $n_{int}$}
 \psfrag{O}{\footnotesize $\rm{Price}$}
 \includegraphics[width=8.5cm]{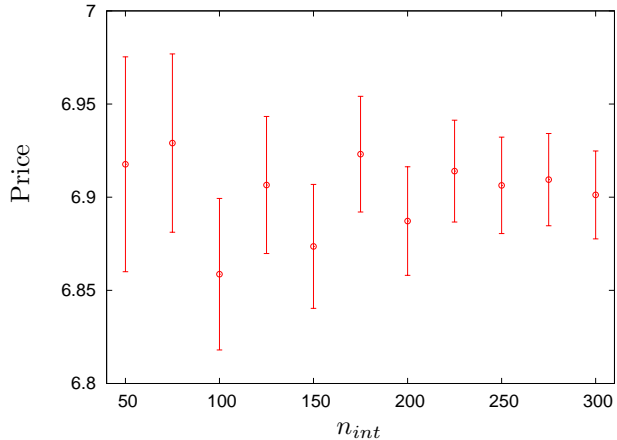}
\end{figure}

To conclude, let us 
comment about the estimate of the numerical error 
for the PITP and the BBST algorithms. Errors in 
Equation~\eref{meanerrorbar} result from the combination 
of the Monte Carlo errors on each end point, $z_{n+1}$. 
To estimate the error associated with the finiteness of $n_{int}$ 
required by the deterministic integration, we 
analyzed the stability of the price with respect to the 
number of integration points. In Figure \ref{f:error}
we show the prices obtained according to this procedure.
 It can be seen that the fluctuations of the price 
value due to the choice of $n_{int}$ become negligible with 
respect to the width of the error bars, as the value of $n_{int}$ 
increases. This is why we consider the relevant 
source of 
numerical error as related to the Monte Carlo part of the integration and 
we set $n_{int}=200$ in our simulations.

\subsubsection{Up-out barrier options\\}\label{s:barrier}

In this section we consider barrier options of European style, 
i.e.~whose exercise is possible only at the maturity.
In particular, we price so called knock-out up options. 
The payoff at maturity $T_{n+1}=T$ is a functional of the 
whole path $\{S(s);0\leq s\leq T\}$ and has the following expression
\beq\label{upout}
 f_U[S]=\rme^{-rT}\max(S(T)-K,0)\Indic_{\tau>T}+\rme^{-r(\tau-t)}R 
 \Indic_{\tau\leq T},
\eeq 	
where $U$ is the upper barrier, $\tau=\inf\{s>0 : S(s)\geq U)$, $R$ is a fixed cash rebate and $\Indic_{\mathcal{A}}$ is the characteristic
function of the set $\mathcal{A}$. In our simulation we set $R=0$. 
It is known~\cite{Lapeyre} that, 
whenever we discretize this continuous time problem, setting 
\beq
 f_U(Z_T,\ldots,z_0)=
 \rme^{-rT}\max(\rme^{Z(T)}-K,0)\prod_{i=0}^{n+1}\Indic_{Z(T_i)<\log U},
\eeq 
we overestimate the price of up-out options. Actually, 
we do not take into account the possibility that the asset 
price could have crossed the barrier for some  
$t\in]T_i,T_{i+1}[$, $i=0$ to $n$. A strategy to obtain a 
better approximation 
is the following. We check, at each 
time step $T_i=i\Delta t$, and for all Monte Carlo paths, 
whether $\exp\{Z^{(l)}(T_i)\}$ has reached the 
barrier $U$. If not, we compute the value
\beq
 p_i\doteq\exp\left[-\frac{2}{\sigma^2 \Delta t}
 (\log U-Z^{(l)}(T_{i-1}))(\log U-Z^{(l)}(T_i))\right],
\eeq  
and we extract a random variable from a Bernoulli distribution with parameter 
$p_i$. If the result is 1, the barrier value has been
reached in the interval $]T_{i-1},T_i[$ and the price associated to  that 
given path is zero. Otherwise, the simulation is carried on. 
This technique is widely discussed in \cite{Lapeyre, Baldi, Gobet}. \par
%
%
\Table{\label{t:upout} Numerical values for the price of up-out 
barrier call options obtained through different algorithms for the
parameters $S_0=100$, 
$r=0.095$, $\sigma=0.2$, $T=1$ year and $n+1=100$. 
Errors correspond to one standard deviation.}
\br
& \centre{4}{K = 100} & \centre{4}{K = 130}\\
\ns
&\crule{4}&\crule{4}\\
& \centre{2}{U = 150} & \centre{2}{U = 200} & \centre{2}{U = 150} & \centre{2}{U = 200}\\
\ns
&\crule{2}&\crule{2}&\crule{2}&\crule{2}\\
& Price & Error & Price & Error & Price & Error & Price & Error\\
\mr
AT-MCRW&9.087&0.012&12.853&0.015&0.647&0.004&2.353&0.011\\
AT-PITP&9.088&0.008&12.830&0.001&0.638&0.002&2.336&0.001\\
AT-PICH&9.099&0.016&12.815&0.014&0.647&0.002&2.333&0.003\\
\br
\end{tabular}
\end{indented}
\end{table}
%
%

In Table~\ref{t:upout} we report prices, with 
the corresponding numerical errors. We price an ATM option, $K=100$, and an 
OTM option, $K=130$, with $U=150$ and $U=200$, all the other
parameters as in Section~\ref{s:asianunidim}. It is
particularly evident that the precision of PITP exceeds that of MCRW, 
both for ATM and OTM options.

\subsubsection{Reverse cliquet options\\}\label{s:reverse}

The last one-dimensional example we consider is 
represented by the reverse cliquet option, 
whose payoff function is 
\beq\label{revcliq}
\fl f_{RC}(Z(T_{n+1}),\ldots,z_0)=
\mathrm{e}^{-rT}\max\left[F,C+\sum_{i=0}^n\min
\left(\frac{S(T_{i+1})-S(T_i)}{S(T_i)},0\right)\right]_{S(T_i)=\exp\{Z(T_i)\}}.
\eeq 
The option is characterized by the number of periods $n$, 
the minimum amount (the \emph{floor}) $F$, and the maximum payable 
coupon (the \emph{cap}) $C$. This option is especially valuable when 
positive performances are more probable.

We have tested our algorithms by choosing the parameters' values 
as in  \cite{Airoldi}, whose 
numerical results for option prices 
are reported in Table \ref{t:rcliquet}, for 
the sake of comparison. The floor $F$ has been fixed equal to zero, 
while the cap $C$ is set equal to $n\,c$, with $c=0.04$. 
The spot is $S_0=100$, $r=0.09$ and $\sigma=0.3$. We also change 
the maturity $T$ by fixing $T/n=1/12$ year and letting $n=4,12,24,36$. 
In Table \ref{t:rcliquet} we show the values corresponding to different 
values of $n$. Integration on $z_{n+1}$ for the 
AT-PITP algorithm is performed on
 a symmetric interval 
 of width $8\sigma\sqrt{T}$ centred on $Z_0+(r-\sigma^2/2)T$.  
 Once again we observe 
 accurate 
 path integral pricing and 
 a good agreement between the results of the various algorithms.   
%
%
\Table{\label{t:rcliquet} Reverse cliquet option fair price 
for $S_0=100$, $r=0.09$, $\sigma=0.3$ and $T/n=1/12$ year. 
Errors are not quoted in \cite{Airoldi}.}
\br
\ns
&\centre{2}{n = 4} & \centre{2}{n = 12} & \centre{2}{n = 24} & \centre{2}{n = 36}\\
\ns
&\crule{2}&\crule{2}&\crule{2}&\crule{2}\\
& Price & Error & Price & Error & Price & Error & Price & Error\\
\mr
AT-MCRW&0.0574&0.0001&0.1223&0.0001&0.1993&0.0002&0.2611&0.0002\\
AT-PITP&0.0572&0.0001&0.1225&0.0002&0.1992&0.0003&0.2611&0.0003\\
AT-PICH&0.0572&0.0001&0.1218&0.0002&0.1990&0.0002&0.2606&0.0003\\
 \cite{Airoldi} &0.0574&  &0.1222&  &0.1990&  &0.2609& \\
\br
\end{tabular}
\end{indented}
\end{table}

%
%
    
 
\subsection{Multidimensional assets}\label{s:basket}

In this section we study the performances of path 
integral pricing in the case of options on multidimensional 
assets $S=(S_1,S_2,\ldots,S_D)$. 
As an example, we price an Asian call option on the basket 
$X$ whose value at time $t$ is obtained by linearly 
combining the values of  the components of $S$:
\begin{equation}
\begin{array}{rcl}
X_{t}&=&\sum_{i=1}^{D}\alpha_{i}S^{i}_{t}\\
\sum_{i=1}^{D}\alpha_{i}&=&1\\
\alpha_{i}>0&&\forall i=1\;{\rm to}\;D .
\end{array}
\end{equation}
Consequently, we have 
\beq
 f_{AD}(Z(T),\ldots,Z_0)=\rme^{-rT} \max\left(\frac{1}{n+2}\sum_{j=0}^{n+1}
 \sum_{i=1}^{D}\alpha_{i}\exp\{Z^{i}(T_j)\}-K,0\right).
\eeq

In order to compare the multidimensional performance of all 
the algorithms introduced in Section \ref{s:algo}, 
we need to choose $D$ such that it still makes sense  to 
perform a deterministic integration over $Z(T_{n+1})=\log S(T)$. 
However, we expect a gain in competitiveness of pure Monte 
Carlo integration (PIFL, PICH), as 
the deterministic approach gradually loses its attractive features 
as the dimension increases. 
That is why we choose a three-dimensional correlated asset.
 
All the tests are performed by setting $r=0.095$ and considering a  
maturity of $T=1$ year with a time discretization of 100 time steps 
($n+1=100$). Unless otherwise specified, $\rho_{i,j}=0.6$ for 
any $i\neq j$, and $\sigma_{k}=0.2$ for all  $k=1$ to $D=3$. 
Path integral integration is limited to the compact subset $\mathcal{C}$ 
described in Section~\ref{s:listofalgo}. In the special case of PITP, 
we consider 1000 Monte Carlo samples for each end point. 
It is worth 
noticing that, if we choose $n_1$ integration values for each dimension,  
the total number of required deterministic 
integration points grows exponentially as $n_1^D$. Thus, 
an apparently poor-quality unidimensional integration with $n_1=10$ 
consists indeed in evaluating the integrand function on $10^3$ points.

The first test concerns the convergence of the deterministic integration. 
We set $K=120$, $S_0=(100,90,105)$ and we compare 
prices obtained with $n_1=6,8,10$, i.e.~with $216\cdot 10^3$, $512\cdot 10^3$, 
$10^6$ total integration points for $z_{n+1}.$
In Table \ref{t:intptsmultidim} we 
show the prices thus obtained with their one standard deviation 
errors, together with the ratio between one standard deviation Monte Carlo 
error and the price 
corresponding to $n_1=10$ (fourth column), and the 
percentage difference between Price$(n_1)$ and Price($n_1=10$) (fifth column). As 
in the unidimensional case, changes in prices due to the number 
of deterministic integration 
points fall well inside the Monte Carlo $95\%$ confidence 
interval $[\mbox{Price}-1.96~\mbox{Error},
\mbox{Price}+1.96~\mbox{Error}]$.
\Table{\label{t:intptsmultidim} Prices, one 
standard deviation Monte Carlo errors, 
percentage errors and prices' percentage  differences 
of an Asian basket call option, according to the 
multidimensional PITP algorithm, with $K=120$. 
The reference for prices' percentage differences is Price($n_1=10$).}
\br
\ns
$n_1$&Price&Error& Relative & Percentage Price\\ 
&&&Error&Difference\\
\mr
10&0.306     &   0.003& 0.99\%& --\\
8&0.310  &      0.005&1.48\%&1.23\%\\
6&0.323   &    0.008&2.38\%&5.9\%\\
\br
\end{tabular}
\end{indented}
\end{table}

As a second test,  we studied the drawbacks of performing  
pure Monte Carlo (PIFL, PICH, MCRW) simulations  with a small 
number of Monte Carlo paths ($10^4$)  finding that in these cases 
MCRW works best, since the path integral fails to efficiently explore 
the support of the integrand function. 
Results corresponding to $\rho_{i,j}=0.8$ for any $i\neq j$, $K=110$, 
$\sigma_k=0.2+0.02\cdot(k-1)$  and obtained with two different  choices 
for the centre $\bar z$
of the integration hyper-cube $\mathcal{C}$ and two different spot $S_0$ 
are reported in Table \ref{t:smallstatistics}. The case $S_0=(100,95,80)$ 
corresponds to an OTM 
option, while when $S_0=(107,109,114)$, corresponds to an ATM option. 
Convergence results 
are poor as the estimated price depends on the integration region and the 
Monte Carlo error is large.
\Table{\label{t:smallstatistics} Prices and errors for Asian 
basket call options according to the algorithms 
PIFL, PICH, MCRW with a small set ($10^4$) of 
Monte Carlo samples. Errors 
correspond one standard deviation. $K=110$}
\br
\ns
&\centre{4}{ATM {\scriptsize ($S_0=(107,109,114)$) }} & 
&\centre{2}{OTM {\scriptsize ($S_0=(100,95,80)$)}}\\
\ns
&\crule{2}&\crule{2}&\crule{2}&\crule{2}\\
&\centre{2}{ $\rme^{\bar z}$=(105,110,115) }&\centre{2}
{ $\rme^{\bar z}$=(100,120,115) }&\centre{2}{ $\rme^{\bar z}$=(110,110,110) }
&\centre{2}{ $\rme^{\bar z}$=(120,120,110) }\\
\ns
&\crule{2}&\crule{2}&\crule{2}&\crule{2}\\
& Price & Error & Price & Error& Price & Error & Price & Error\\
\mr
MCRW&  7.5&0.1&7.5&0.1&0.83&0.03&0.83&0.03\\
PIFL&  6.3&0.5&7.8&0.6&0.77&0.09&0.78&0.08\\
PICH&  6.8&0.4&8.1&0.4&0.80&0.09&0.56&0.07\\
\br
\end{tabular}
\end{indented}
\end{table}

Once we take care of choosing a sufficient number of Monte 
Carlo paths \footnote{Just take as reference the deterministic 
integration with the 
rule of thumb of setting at least 6 integration
 points for each dimension and 1000 MC paths for each end
  point, i.e.~$6^D\cdot 10^3$ total samples.}, we compare the 
performance of our algorithms in the cases of ATM and OTM options. 
We report the results of such analysis 
in Table \ref{t:goodstatistics}, where $S_0=(100,90,105)$ and the strike 
$K$ is varied to perform 
ITM/ATM ($K=100$) and OTM ($K=140$) pricing.  We report 
results obtained with two differently centred  integration 
intervals, in order to show that the chosen number of samples 
is enough to guarantee stability of integration to (relatively small) 
changes of the integration hypercube\footnote{It is important to recall that, 
if we perform integration on an interval 
whose spot values are too low, we will have an under-estimation of the price.}.   
\Table{\label{t:goodstatistics} Prices and errors for Asian basket call 
options according to the algorithms PITP, PIFL, PICH, MCRW 
with $n_1=6$ and 216000 total Monte Carlo paths. $S_0=(100,90,105)$. 
Errors correspond to one standard deviation.}
\br
\ns
&\centre{4}{ATM {\scriptsize ($K=100$)}} & 
&\centre{2}{OTM {\scriptsize ($K=140$)}}\\
\ns
&\crule{2}&\crule{2}&\crule{2}&\crule{2}\\
&\centre{2}{ $\rme^{\bar z}$=(110,100,110) }&\centre{2}
{ $\rme^{\bar z}$=(100,100,100) }&\centre{2}{ $\rme^{\bar z}$=(140,140,140) }
&\centre{2}{ $\rme^{\bar z}$=(130,130,130) }\\
\ns
&\crule{2}&\crule{2}&\crule{2}&\crule{2}\\
& Price & Error & Price & Error& Price & Error & Price & Error\\
\mr
MCRW&5.29&0.02&5.29&0.02&0.0049&0.0004&0.0049&0.0004\\
PITP&5.33&0.04&5.28&0.04&0.0051&0.0003&0.0049&0.0003\\
PIFL&5.37&0.06&5.41&0.07&0.0048&0.0003&0.0048&0.0002\\
PICH&5.26&0.03&5.28&0.03&0.0048&0.0001&0.0050&0.0001\\
\br
\end{tabular}
\end{indented}
\end{table}
When compared to Table~\ref{asiantable}, the results 
shown in Table~\ref{t:goodstatistics} present some 
similarities and some differences. 
As in the unidimensional case, the path integral is still a valuable choice 
to price OTM options, prices 
being in agreement with the benchmark MCRW and errors smaller, 
especially with a Cauchy pdf sampling (PICH). As expected, however,
 the external deterministic integration (PITP) has effectively lost 
 its attractive properties, PIFL and PICH giving more precise 
 confidence intervals. Let us stress that, when the dimensionality increases, 
 the path integral performance for ATM options worsen. 
 Even if we perform  tests with 
 $5\cdot 10^5$  and $10^6$ Monte Carlo samples, 
 whenever we price ITM/ATM options, the path integral method is a bit 
 less precise than the random walk, the central value 
 being nevertheless compatible with the benchmark results. 
 Because the performances for OTM options are good, 
 we infer that, to increase the accuracy in pricing 
 high dimensional ITM/ATM options, we should consider 
 an hyper-cube $\mathcal{C}$ wider than four standard deviation. 


\section{Greeks}\label{s:greek}

In the present section, 
we report our results concerning the computation of 
the Greeks for unidimensional Asian and barrier 
knock out options. Actually, 
it is interesting to compare the numerically estimated 
exotic Greeks with those implied by the Black and Scholes model for 
plain vanilla call options.
The values of the parameters are $K=100$, $r=0.095$, $\sigma=0.2$, 
$T=1$ year 
and $n+1=100$, while for the barrier we choose $U=150$. 

\begin{figure}[ht]\caption{\label{f:greek} Option price 
and Greeks for plain vanilla (\full), Asian (\opensquare) 
and barrier knock out (\opencircle) call
 options. }
\begin{minipage}[b]{0.55\textwidth}
 \centering
 \psfrag{S}[b]{\footnotesize $S$}
 \psfrag{Option Price}{\footnotesize Option Price}
 \includegraphics[width=8.5cm]{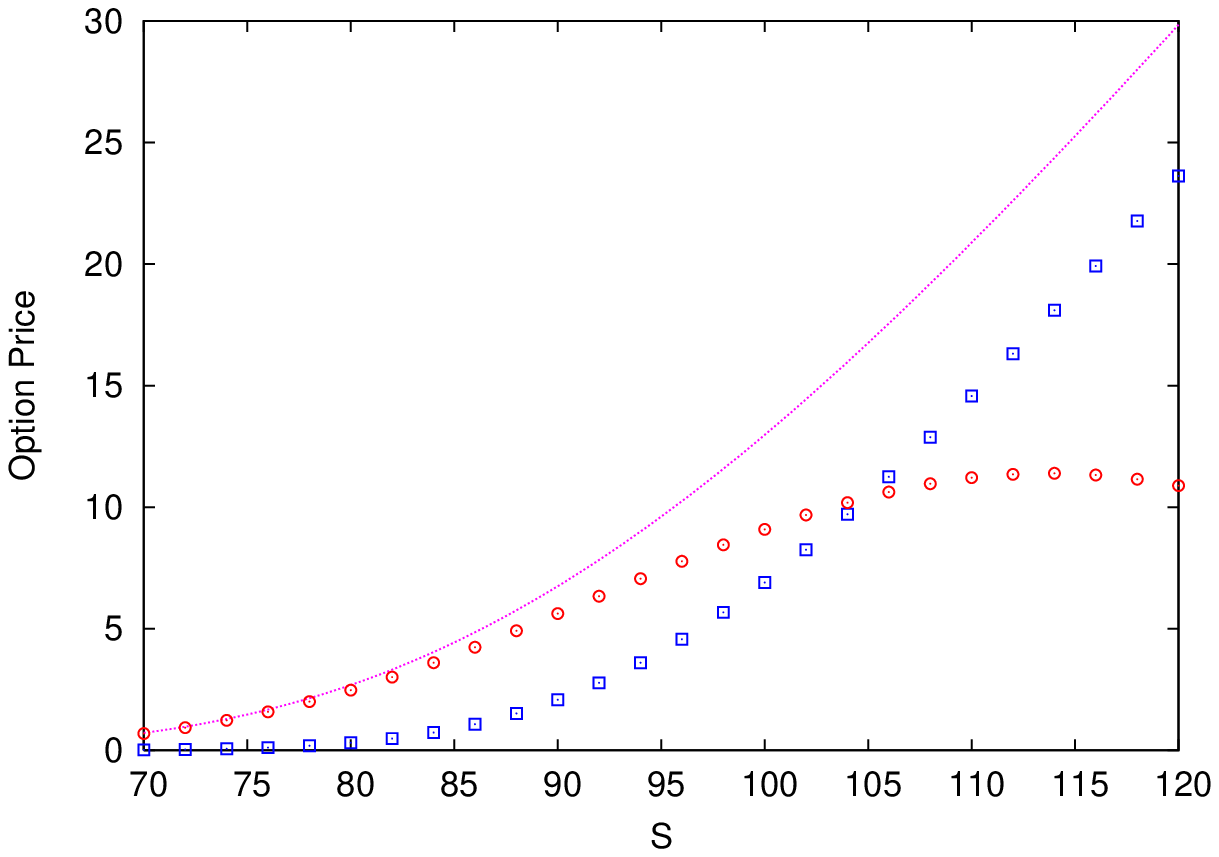}
\end{minipage}
\begin{minipage}[b]{0.5\textwidth}
 \centering
 \psfrag{S}[b]{\footnotesize $S$}
 \psfrag{Delta}{\footnotesize $\Delta\doteq\partial O/\partial S$}
 \includegraphics[width=8.5cm]{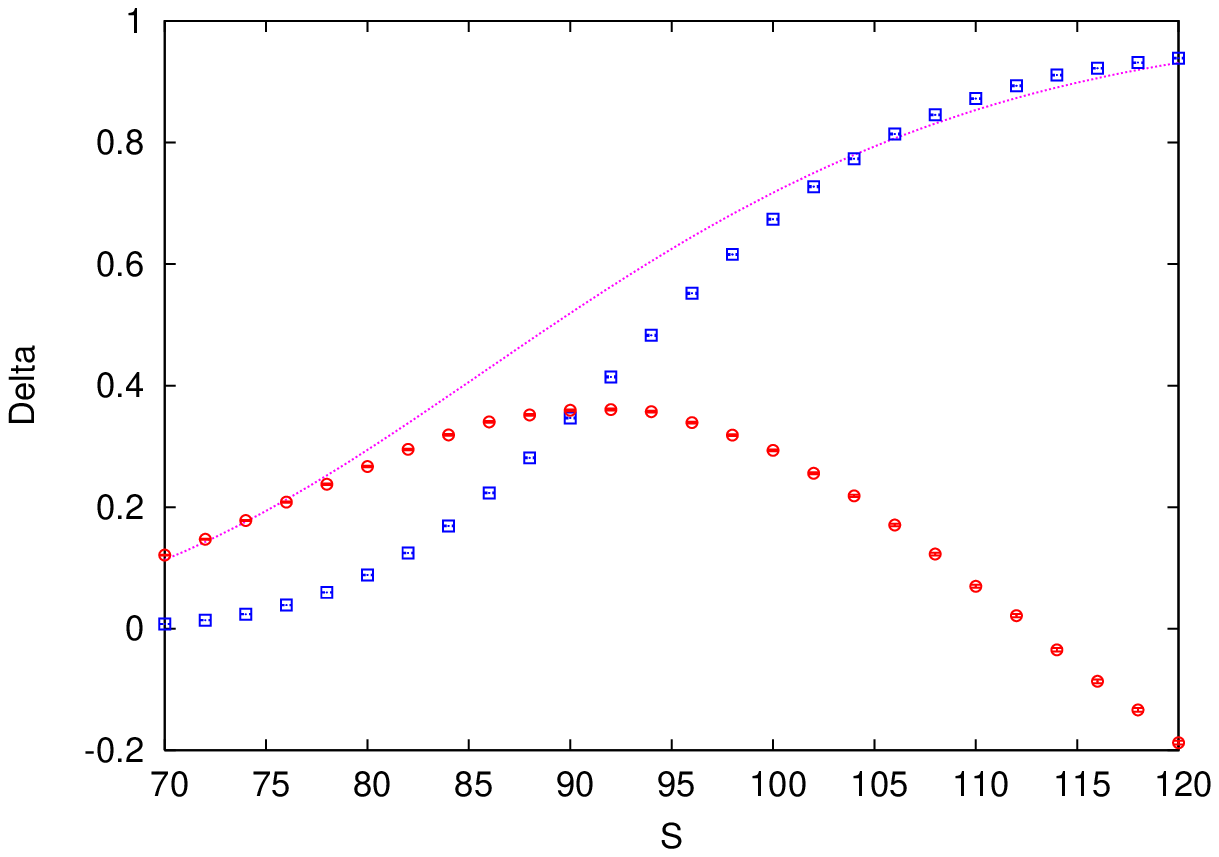}
\end{minipage}
\end{figure}

In Figure~\ref{f:greek} we show the price of the option and the 
Greeks delta, gamma, vega and theta as functions of the spot price $S$.
In our approach the Greeks are numerically computed using a 
finite difference
method applied to the option values returned by the 
path integral with trapezoidal integration. The error 
bars are obtained by propagating the (one standard deviation) numerical error
of the path integral option prices. Therefore, the values of the Greeks are
more accurate than those obtained with other approaches when the path integral
pricing outperforms existing techniques.

\begin{figure}[ht]
\begin{minipage}[b]{0.55\textwidth}
 \centering
 \psfrag{S}[b]{\footnotesize $S$}
 \psfrag{Gamma}{\footnotesize $\Gamma\doteq\partial^2 O/\partial S^2$}
 \includegraphics[width=8.5cm]{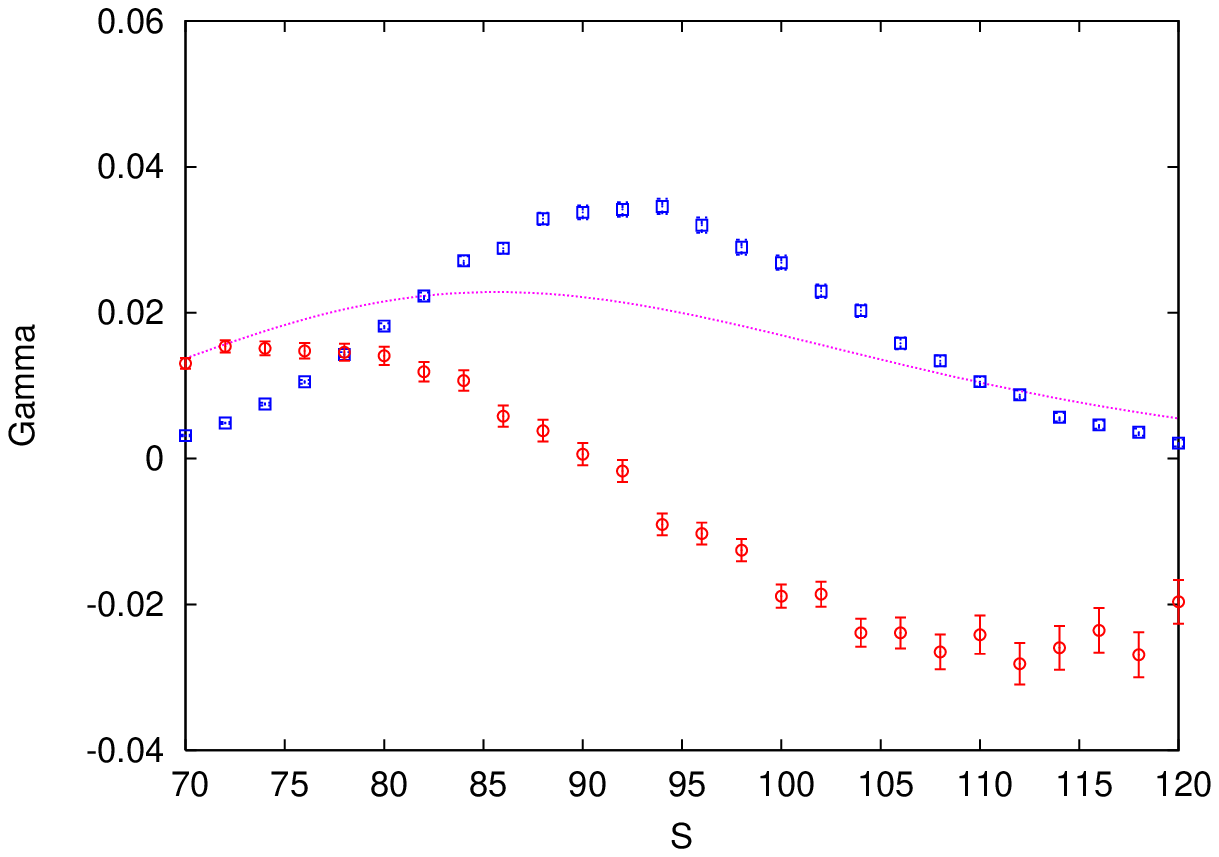}
\end{minipage}
\begin{minipage}[b]{0.5\textwidth}
 \centering
 \psfrag{S}[b]{\footnotesize $S$}
 \psfrag{Vega}{\footnotesize $V\doteq\partial O/\partial\sigma$}
 \includegraphics[width=8.5cm]{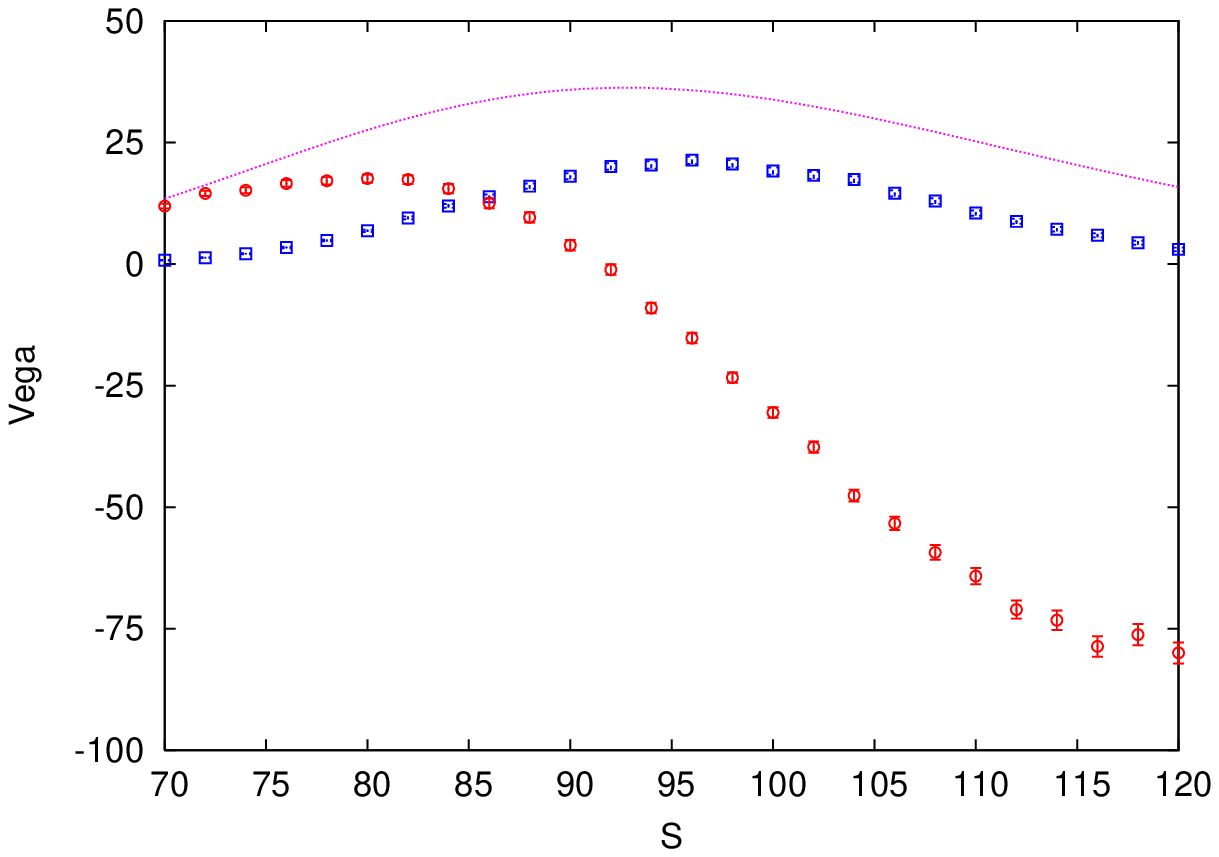}
\end{minipage}
\end{figure}

\begin{figure}[ht]
\begin{minipage}[b]{0.55\textwidth}
 \centering
 \psfrag{S}[b]{\footnotesize $S$}
 \psfrag{Theta}{\footnotesize $\Theta\doteq-\partial O/\partial T$}
 \includegraphics[width=8.5cm]{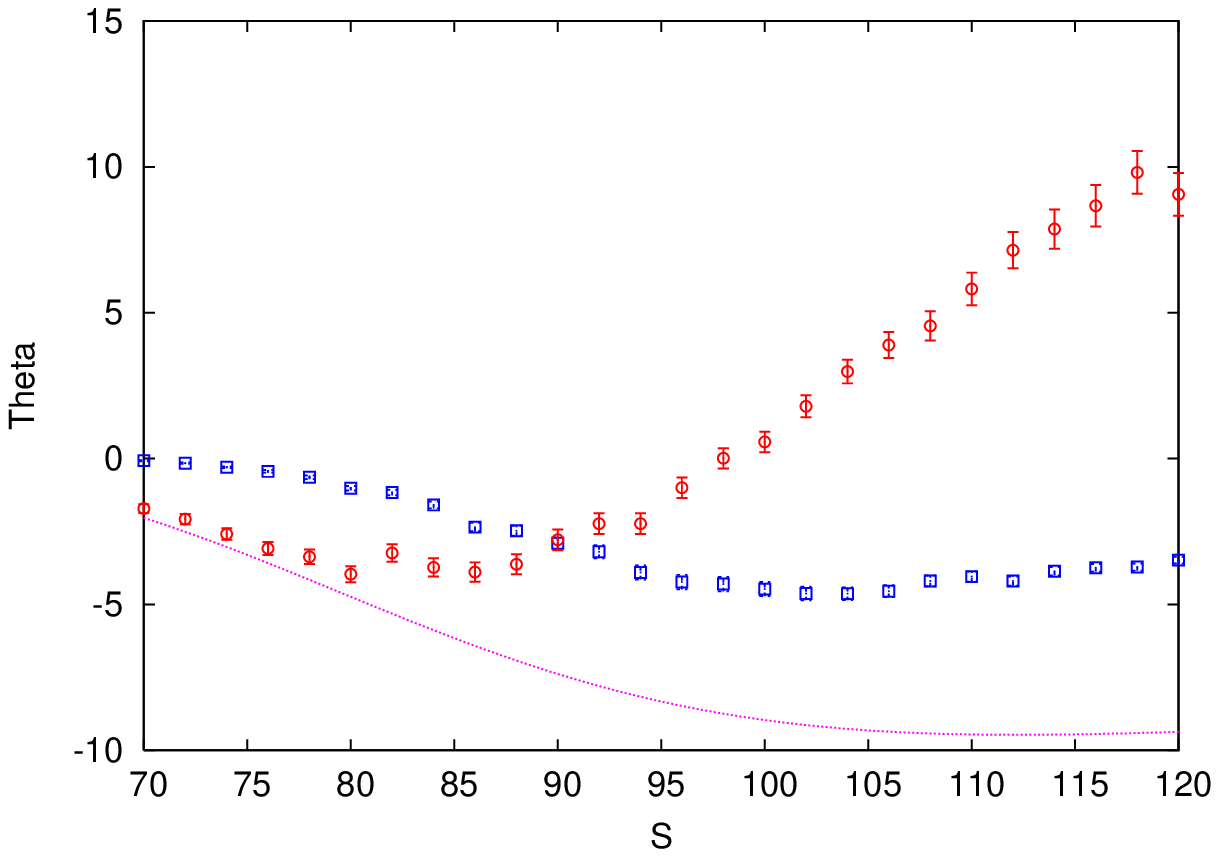}
\end{minipage} 
\begin{minipage}[b]{0.5\textwidth}
\end{minipage}
\end{figure}

As expected, the qualitative behaviour of prices 
and Greeks for Asian call options 
does not differ significantly from that of plain vanilla ones. 
The shift in prices is due to the fact that in the Asian payoff the role
of $S(T)$ is played by the mean value of $S$ along the path. 
A lower price is therefore a straightforward consequence
of the fact that we have $ \mathbb{E}[S(T_1)+\cdots
S(T_{n+1})]/(n+1)\leq \mathbb{E}[S(T_{n+1})]$. 
The Greeks do not coincide exactly, but the relevant features, 
such as the sign of the derivative, are preserved. 
The situation is completely different for the barrier options. 
First of all, we expect that for small values of $S$ 
and with our choice for the max barrier value, $U=150$, the results 
overlap the European ones. This can be easily verified 
from Figure~\ref{f:greek}  and considered as a check 
of the consistency of our numerical results. 
The profile of the
price graph is characterized by changes both of the monotonic properties 
and of the concavity, as shown in Figure~\ref{f:greek}. 
This reflects in the
change of sign of delta and gamma. The behaviour 
of vega can be explained by
noticing that, for $S\ll U$, an increase of the $\sigma$ means 
an increase in the width of the distribution of $S(T)$, that 
reaches higher
values without reaching the barrier, 
so $\partial{O}/\partial \sigma>0$. Conversely, 
when $S$ and $\sigma$ grow, $S(T)$ 
reaches the barrier more frequently and the option loses
value. Analogous reasoning applies to theta, 
i.e.~$-\partial{O}/\partial T$, where the role of $\sigma$ 
is played by the maturity $T$.
However, in this case, the presence of the minus sign in the 
definition implies $\Theta<0$ for $S\ll U$ and $\Theta>0$ otherwise.

\section{Conclusions and outlook}\label{s:conclusion}

In this paper we have shown how the path integral approach 
to stochastic processes can be
successfully applied to the problem of pricing exotic derivative contracts. 
Numerical results
for the fair price and the Greeks of a variety of options have 
been presented and
compared with those obtained by means of other standard 
(such as the Monte Carlo Random Walk) and non standard 
(see \cite{Airoldi}) approaches employed in quantitative finance. 
With respect to the original formulation of 
 \cite{Montagna}, the method has been generalized
in order to cope  with options depending on multiple and correlated 
underlying assets.
Concerning options depending on a single asset, it has been shown 
that the algorithm can provide 
very precise results, especially when pricing ATM and OTM options. 
This is due to an appropriate
separation of the integrals entering the path integral pricing formula and, 
more importantly, to a 
careful simulation of random paths with fixed end points, able to
probe the regions contributing to the dominant part of the payoff functions.  
As far as (multidimensional) basket options are concerned,
while the standard Monte Carlo simulation turns out to be 
more efficient for ITM/ATM options, our 
approach exhibits better performances for OTM options. 

In all the cases, the computational time is essentially the same required by a 
standard Monte Carlo calculation.
The algorithm is general and could be extended to 
price other types of exotic contracts.

As a future important development, it would be interesting to explore the feasibility 
of an application of our framework to derivative pricing approaches beyond 
Black-Scholes and dealing with  non-Gaussian features of financial time series, 
such as Bouchaud-Sornette residual risk minimization \cite{Bouchaud0,Bouchaud1,Matacz}, 
stochastic volatility models \cite{Heston0,Fouque,Perello},
multi fractal random walks \cite{Bacry,Pochart}, GARCH processes \cite{Heston1}
and other generalizations of the log-normal dynamics \cite{Aurell,Borland,Kleinert}.


\ack

We would like to thank Bernard Lapeyre for suggesting to us that we compare 
with the Brownian bridge and stratification technique, 
as well as two anonymous referees for some pertinent and useful 
remarks concerning the layout and structure of the paper. 
We acknowledge partial collaboration with Francesca Rossi at the 
early stage of this work. 
We wish to thank Carlo Carloni Calame for helpful assistance with 
software installation. The work of G.B. is partially supported by 
STMicroelectronics. 


\appendix

\section{Stratification and Brownian bridge}\label{s:bbridge}

We describe here  the algorithm used to test our hints about the 
reasons of the good performances of path integral 
with deterministic integration when pricing path dependent options 
on unidimensional assets. In order to improve the numerical precision, 
it is necessary to lead Monte Carlo paths  to a region in 
which the payoff function is different from zero. 
This is the advantage of performing the external integral 
in Equation~\eref{targetIO} in a clever way and to drive 
paths toward some fixed end points.
The algorithm here described exploits this idea 
by means of a backward construction of the underlying process 
(Brownian bridge), instead 
of using a path integral method. For simplicity, since this method 
has only been used in the case $D=1$, we outline here 
the construction of a unidimensional Brownian bridge only.

The aim is to describe the law of $Z(t)$, with $t\in [0,t']$, once 
$Z_0$ and $Z(t')$ are known.  By setting  
\beq
\label{backpath} Z(t)=Z_0+\frac{Z(t')-Z_0}{t'}t+\varepsilon(t),
\eeq
it is easy to see that $\varepsilon({t})$ is Gaussian with zero mean and  
variance $t(t'-t)\sigma^{2}/t'$. It is worth 
noticing that, while  
Law$[Z(t)|Z(t')]\neq $Law$Z(t)$, $\varepsilon(t)$ is independent of $Z(t')$.
Hence, given $Z_0$ and $Z(T)$, we construct  Monte Carlo paths 
$Z_0,Z(T_1),\ldots,Z(T)$ by recursively applying 
Equation~\eref{backpath} from $t'=T,\;t=T_{n}$ down to $t'=T_2,\; t=T_1$. As 
for the path integral framework of Equation~\eref{logPath} and 
the standard Euler 
discretization scheme of Equation~\eref{logEuler}, sampling a path is  
equivalent to generating i.i.d.~standardized Gaussian variables. 
The main difference with the path integral method is that here, once we
 have the $l$-th  Gaussian sample 
$\{\varepsilon^{(l)}(T_1),\ldots ,\varepsilon^{(l)}(T_n)\}$, 
we are nevertheless forced to simulate the path \emph{recursively}, which
means that  
we cannot infer $Z^{(l)}(T_i)$ if we have not sampled $Z^{(l)}(T_{i+1})$. 
In the path integral case, instead, extraction is straightforward 
by means of Equation~\eref{logPath}.
   
Let us now explain how to compute Equation~\eref{targetExpectation} 
by means of the Brownian bridge. 
By setting 
$\bar f(z_{n+1},y_{n}\ldots,y_{1},z_0)\doteq f(z_{n+1},z_n\ldots,z_0)$ with 
the replacements 
\begin{equation*}
z_i=z_0+\frac{z_{i+1}-z_0}{T_{i+1}}T_i+y_i,\quad i=1,\ldots,n,
\end{equation*}
we have 
\begin{eqnarray}\label{targetBBST}
\fl\mathbb{E}[f(Z(T_{n+1}),\ldots,Z_0)]=\mathbb{E}
[\bar f(Z(T_{n+1}),\varepsilon(T_n),\ldots,\varepsilon(T_1),z_0)]\nonumber\\
=\int_{\mathbb{R}^{D}}\!\!\rmd z_{n+1}\rho_{n+1}(z_{{n+1}})
\int_{\mathbb{R}^{D\times n}}\prod_{i=1}^{n}\left(\rmd y_{i}
\rho_{\varepsilon,i}(y_{i})\right)\bar f(z_{n+1},y_n,\ldots,y_1,z_0)\\
=\int_{\mathbb{R}^{D}}\!\!\rmd z_{n+1}\rho_{n+1}(z_{{n+1}})
\mathbb{E}[\bar f(z_{n+1},\varepsilon(T_n),\ldots,\varepsilon(T_1),z_0)],
\nonumber
\end{eqnarray}
where $\rho_{n+1}$ is the pdf of $Z(T)$ and $\rho_{\varepsilon,i}$ 
is the pdf of $\varepsilon(T_i)$, for $i=1$ to $n$.
  
It is now straightforward to see that Equation~\eref{targetBBST} has 
the same form as Equation~\eref{targetIO}. Therefore, we can 
approximate integration over $z_{n+1}$ as in 
Equation~\eref{GQuad} and evaluate {\it via} Monte Carlo the inner 
mathematical expectation.

By means of Equation~\eref{GQuad}, we are, 
in some sense, stratifying 
the domain $\mathcal{D}$ of the random variable $Z(T_{n+1})$ by 
dividing it into disjoint sub-sets $\mathcal{D}_i$ (here the 
sub-sets reduce to the points $z^{i}_{n+1}\in\mathbb{R}$ of 
Equation~\eref{GQuad}).  For each sub-set, we then compute the inner 
integral by forcing $Z(T_{n+1})\in\mathcal{D}_i$. 
It is possible to show \cite{Lapeyre} that this procedure 
may lead to variance reduction. 
This way of proceeding has the same qualities and the same 
limitations as the PITP, yielding less accurate results
for multidimensional assets.  


\end{document}